\documentclass[twocolumn,showpacs,preprintnumbers, 
               superscriptaddress,
               eqsecnum,nofootinbib]{revtex4}
\usepackage{graphicx, fancybox}
\usepackage{amsmath,amssymb}
\usepackage[dvipdfm, colorlinks=true, pdfstartview=FitV, linkcolor=red, citecolor=blue, urlcolor=blue]{hyperref}
\usepackage{color}

\newcommand{\vect}[1]{\mathbf{#1}}

\newcommand{\Slash}[1]{\ooalign{\hfil/\hfil\crcr$#1$}}

\newcommand{\vp}{\vect{p}}
\newcommand{\vk}{\vect{k}}

\newcommand{\vx}{\vect{x}}
\newcommand{\vgamma}{{\boldsymbol \gamma}}

\newcommand{\cp}{g}
\newcommand{\Tr}{\mathrm{Tr}}

\newcommand{\comment}[1]{}

\newcommand{\nf}{n_F}
\newcommand{\nb}{n_B}
\newcommand{\Nc}{N_c}
\newcommand{\Nf}{N_f}
\newcommand{\mf}{m_f}

\begin{document}

\title{Chiral symmetry breaking and confinement effects on \\dilepton and photon production around {\bf $T_c$}
}

\author{Daisuke Satow}
\email{sato@fbk.eu}
\affiliation{ECT*, Villa Tambosi, I-38123 Villazzano (Trento), Italy}

\author{Wolfram Weise}
\email{weise@tum.de}
\affiliation{ECT*, Villa Tambosi, I-38123 Villazzano (Trento), Italy}
\affiliation{Physik-Department, Technische Universit\"{a}t M\"{u}nchen, D-85747 Garching, Germany}

\begin{abstract} 
Production rates of dileptons and photons from the quark-gluon (QGP) phase are calculated taking into account effects of confinement and spontaneous chiral symmetry breaking ($\chi$SB) not far from the transition temperature $T_c$.
We find that the production rates of dileptons with large momenta and of photons originating from the QGP around $T_c$ are suppressed by the  $\chi$SB effect. We also discuss to what extent information about details of the chiral transition, such as its characteristic temperature range and the steepness of the crossover, are reflected in these quantities.
\end{abstract} 

\date{\today}


\maketitle

\section{Introduction}
\label{sec:intro}

In the limit of massless quarks with $\Nf$ flavours, quantum chromodynamics (QCD) has an exact chiral $SU(\Nf)_L\times SU(\Nf)_R$ symmetry.
Dynamical quark mass generation implies that this symmetry is spontaneously broken.
In addition, non-zero current quark masses in the QCD Lagrangian break chiral symmetry explicitly.
Spontaneous chiral symmetry breaking ($\chi$SB) is manifest in the temperature dependence of the chiral (quark) condensate, $\langle \overline{q}q\rangle$. 
Recent determinations of $\langle \overline{q}q\rangle_T$ from lattice QCD with $\Nf=2+1$ flavours and physical current quark masses display a chiral crossover at a transition temperature around $T_c\simeq 155$ MeV\footnote{This updated value of $T_c$, slightly lowered from a previous determination~\cite{Bazavov:2009zn}, resulted from improvements in controlling the continuum limit.
}~\cite{Bazavov:2011nk, Borsanyi:2010cj}.
Schematic models of the Nambu \& Jona-Lasino (NJL) type~\cite{Nambu:1961tp} link the dynamically generated quark masses, $m(T)$, to the chiral condensate:
\begin{align}
\frac{m(T)}{m(0)}\sim \frac{\langle \overline{q}q\rangle_T}{\langle \overline{q}q\rangle_0}.
\label{eq:condensate}
\end{align}
This concept is adopted in the present work, as will be explained in the next section.

The other important aspect of QCD at temperatures around and above the transition temperature ($T_c$) is the deconfinement transition.
At zero temperature quarks and gluons are confined inside hadrons. Above the deconfinement temperature the quarks and gluons are released and become active degrees of freedom. 
An order parameter for this transition to the quark-gluon plasma (QGP) is the Polyakov loop. 
In the pure gauge case ($\Nf=0$) deconfinement emerges as a first-order phase transition, whereas it becomes a continuous crossover in the presence of dynamical quarks. Thus, in full QCD, the Polyakov loop is not an order parameter in the strict sense but it nonetheless serves as a monitor for deconfinement as a rapid crossover transition. Motivated by this scenario, models have been designed (referred to as PNJL models)~\cite{Fukushima:2003fw} that unify the Polyakov-loop-induced suppression of color non-singlet degrees of freedom below $T_c$ with the NJL mechanism of spontaneous chiral symmetry breaking.

High-energy heavy-ion collisions (HIC) are the experimental tool for investigating properties of the QGP.  Electromagnetic probes such as dileptons and photons~\cite{Vujanovic:2013jpa,Shen:2013vja,Gale:2012xq,Gale:2014dfa, PHENIX:2012, Turbide:2003si, Schenke:2010nt, Schenke:2010rr} produced in these collisions are particularly important messengers for the physics of the hot and dense QCD matter since they are barely affected by successive interactions on their way through the surrounding expanding medium. 
At the Relativistic Heavy Ion Collider (RHIC), the PHENIX experiment has measured~\cite{PHENIX:2012} a large elliptic anisotropy of the produced photons characterized by the quantity $v_2$.
Since the anisotropy is considered to be produced by the collective flow at later stages of the HIC, it is natural that the produced hadrons, mainly generated at the later stages, have large $v_2$. 
By contrast, photons are generated also at early stages of the HIC (in the initial state~\cite{Aurenche:2006vj}, or in the thermalization process~\cite{McLerran:2014hza}).
It is therefore not easy to explain~\cite{vanHees:2011vb} the large photon $v_2$, and currently no theory has completely succeeded in doing it.
Recently, a new mechanism to generate a large photon $v_2$ was suggested~\cite{Gale:2014dfa}.
It is based on the suppression of the photon production in the QGP phase due to the confinement effect.
Nevertheless, the $\chi$SB effect was not taken into account in this analysis, so if this effect further suppresses the photon production from the QGP phase, one can expect that the photon $v_2$ is more enhanced.

Motivated by these issues, we calculate the production rates of dileptons and photons in the QGP phase, using a model in which the confinement and the $\chi$SB effects are both taken into account.
The present work focuses in this context primarily on the role of the latter effect, in terms of dynamically generated constituent quark masses around $T_c$. 
It is an extension of the previous Refs.~\cite{Gale:2014dfa, DileptonPhoton} in which corresponding studies have been performed with quark masses set equal to zero.

This paper is organized as follows:
The next section focuses on the implementation of the effects of confinement and $\chi$SB in our model. The dilepton production rate is calculated and the $\chi$SB effect on this rate is discussed in Sec.~\ref{sec:dilepton}.
Section~\ref{sec:photon} is devoted to the calculation of the photon production rate, and to the discussion on how the $\chi$SB affects this rate.
For this purpose, the photon production rate in the case of massless quark is also calculated beyond leading-log order\footnote{Leading-log approximation means: regarding a quantity of order one as subleading compared with a quantity of order $\ln(1/\epsilon)$ if $\epsilon$ is very small.
In the case of photon production, $\epsilon=g^2T/E$, where $E$ is the photon energy, $g$ the coupling constant, and $T$ the temperature, respectively.
The production rate in the massless case at leading-log order was already calculated in Refs.~\cite{Gale:2014dfa, DileptonPhoton}.}. 
We summarize and present concluding remarks in Sec.~\ref{sec:summary}.
The detailed derivation of the photon production rate for vanishing quark masses, $m=0$,
is given in the Appendix.

\section{Implementing Confinement and chiral symmetry breaking effects}
\label{sec:}
This section describes the treatment of confinement and spontaneous chiral symmetry breaking within the framework of the present model.

\subsection{Polyakov loop}
\label{ssc:Polyakov-loop}

\begin{figure*}[t] 
\begin{center}
\includegraphics[width=1.00\textwidth]{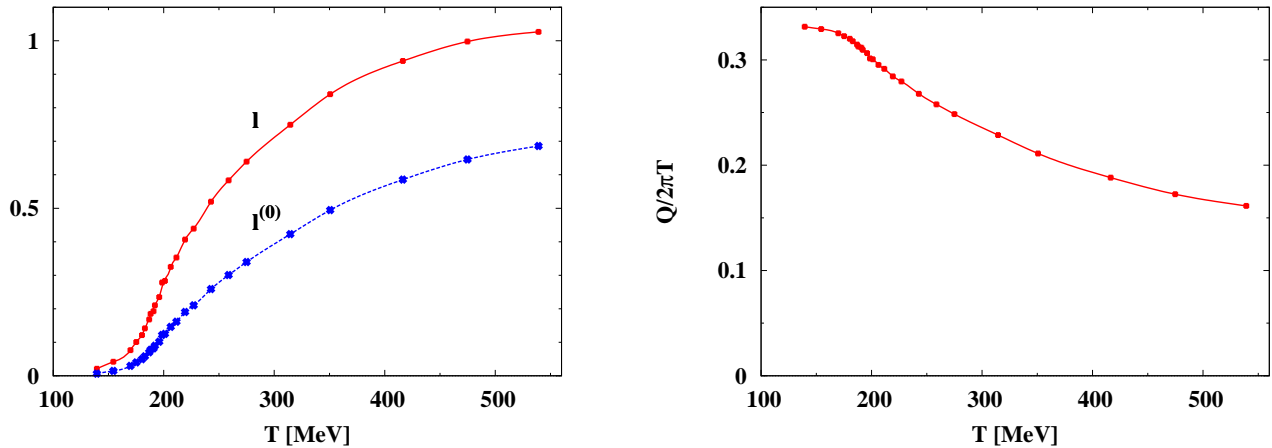}
\caption{Left panel: The Polyakov loop ($l$) deduced from lattice QCD computations~\cite{Bazavov:2009zn}, and the reduced Polyakov loop in which the perturbative correction is removed ($l^{(0)}$), as a function of $T$.
Right panel: temperature dependence of $Q$.
Here $\varLambda_{\overline {\text{MS}}}=T_c/1.35$, where $T_c=155$ MeV~\cite{Bazavov:2011nk, Borsanyi:2010cj}.
The curves in both panels connecting the points are drawn to guide the eye.
}  
\label{fig:Polyakov-Q}
\end{center}
\end{figure*}

The confinement effect is taken into account as a modification of the quark and gluon thermal distribution functions, in a so-called semi-QGP model~\cite{Gale:2014dfa, Hidaka:2008dr, Hidaka:2009hs, Lin:2013efa, DileptonPhoton}. 
The starting point is QCD with $\Nc$ colors and with finite averages of the temporal component of the background gluon field ($A^0$), given by: 
\begin{align} 
A^0&=\frac{i}{\cp}{\text{diag}}(Q^1, Q^2,..., Q^{\Nc}),
\end{align}
where $\cp$ is the coupling.
$A^0$ is assumed to be diagonal in color space and constant in space and time. The spatial components of the background gluon field, $A^i$, are set equal to zero.
The tracelessness of the gauge field implies $\sum_a Q^a=0$. 
Neglecting fluctuations of $A_0$, this quantity is related to the expectation value of the Polyakov loop ($l$) by
\begin{align}
\begin{split}
l&\equiv \frac{1}{\Nc}\Tr L 
= \frac{1}{\Nc}\sum_a e^{i\beta Q^a},
\end{split}
\end{align}
where 
\begin{align}
L(\vx)&\equiv \left\langle{\cal{P}}\exp\left(\int^\beta_0 d\tau g A^0(\tau,\vx)\right) \right\rangle
\end{align} 
 is the Wilson line in the temporal direction in imaginary time, with $\beta\equiv 1/T$, and $\cal P$ denotes path ordering. 

When performing calculations involving quarks and gluons, the background gluon field acts as an imaginary chemical potential coupled to color charges.
The quark and gluon distribution functions, $\nf(k^0)\equiv (e^{\beta k^0}+1)^{-1}$ and $\nb(k^0)\equiv(e^{\beta k^0}-1)^{-1}$, are modified as~\cite{Hidaka:2009hs}
\begin{align}
\nf{}_a(k^0)&= \frac{1}{e^{\beta(k^0-iQ^a)}+1},\\
\nb{}_{ab}(k^0)&=  \frac{1}{e^{\beta(k^0-iQ^{ab})}-1},
\end{align}
where $a$ and $b$ are color indices in the double-line notation~\cite{Hidaka:2009hs, Cvitanovic:1976am}, running from $1$ to $\Nc$, and $Q^{ab}\equiv Q^a-Q^b$.
It might appear that the distribution functions have imaginary parts, but in fact, after summing over color indices, these distributions turn out to be real as they should. To see this, consider the color-averaged quark distribution function: 
\begin{align}
\label{eq:average-quark-distribution}
\langle \nf{}_a(k^0)\rangle &\equiv {1\over N_c} \sum_a \nf{}_a(k^0) 
= \sum^\infty_{n=1} (-1)^{n+1}e^{-\beta k^0 n}\, l_n~,
\end{align}
where we have introduced $l_n\equiv \Tr L^n/{\Nc}$. 
Since $l_n$ is real, it follows that $\langle \nf{}_a(k^0)\rangle$ is real.

Next we demonstrate the suppression of the distribution functions for color-non-singlet degrees of freedom in the confined phase.
From Eq.~(\ref{eq:average-quark-distribution}), the averaged quark distribution function becomes
\begin{align}
\label{eq:nf-confined}
\langle \nf{}_a(k^0)\rangle &= \frac{1}{e^{\beta \Nc k^0}+1},
\end{align} 
where we have used the following expression, valid in the confined phase for odd\footnote{The expression remains valid also in the case of even $\Nc$, but with an unphysical result.
For example, when $\Nc=2$, the baryonic excitations appearing in the confined phase are diquarks, i.e. bosons, whereas the distribution function Eq.~(\ref{eq:nf-confined}) is fermionic.
For this reason, we restrict ourselves to the odd-$\Nc$ case. } $N_c$: 
\begin{align}
\label{eq:ln-confined}
l_n=\left\{ \begin{array}{ll}
(-1)^{k(\Nc+1)} & (n=k\Nc) \\
0 & ({\text{otherwise}}) ~.
\end{array} \right. 
\end{align}
Non-trivial contributions to $l_n$ in the expansion (\ref{eq:average-quark-distribution}) arise whenever $n$ is an integer multiple of (odd) $N_c$. We see that $\langle \nf{}_a(k^0)\rangle$ is suppressed compared to the distribution in the deconfined limit, $Q^a=0$, and it vanishes at large $\Nc$. For $N_c = 3$ one notes that $\langle \nf{}_a(k^0)\rangle$ represents color-singlet 3-quark objects (with energy $E = 3k^0$ and baryon number $B = 1$), although these structures are not localised or clustered in space. Sometimes this property is called ``statistical confinement". The gluon distribution function is likewise suppressed after taking the color sum.

For the $\Nc=3$ case $Q^a$ can be written in the representation 
\begin{align}
\label{eq:Qa-N=3}
(Q^a)= (-Q, 0, Q),
\end{align}
where $Q$ with  $0<Q< 2\pi T$ is the only independent quantity.
The Polyakov loop reads: 
\begin{align}
\label{eq:l-Q}
l=\frac{1+2\cos(\beta Q)}{3}.
\end{align}
In the asymptotic deconfined phase ($l=1$), we have $Q=0$, while in the confined phase, $Q=2\pi T/3$ for the pure glue case ($l=0$). A correspondingly similar expression holds for $l_n$:
\begin{align}
\label{eq:ln-Q}
l_n&=\frac{1+2\cos(n\beta Q)}{3}.
\end{align}

In the present work, $A^0$ and $Q$ are determined from lattice QCD data in the following way~\cite{Lin:2013efa}:
First, we remove the perturbative correction~\cite{Gava:1981qd,Burnier:2009bk} from the Polyakov loop, focusing on nonperturbative effects:
\begin{align}
\label{eq:correction-Polyakov}
l(Q=0)&= 1+\delta l(Q=0),\\
\nonumber
\delta l(Q=0)&= \frac{\cp^2 C_f m_D}{8\pi T}
+\frac{\cp^4 C_f}{(4\pi)^2}
\Bigl[-\frac{N_f}{2}\ln 2\\
&~~~+\Nc\left(\ln \frac{m_D}{T}+\frac{1}{4}\right)\Bigr], 
\end{align}
where $C_f\equiv(\Nc^2-1)/(2\Nc)$ and $m_D$ is the Debye mass of the gluon.
We use the running coupling constant and the expression of the Debye mass estimated by ``fastest apparent convergence'' criteria~\cite{Burnier:2009bk,Kajantie:1997tt}: 
\begin{align} 
\nonumber
g^2&= 24\pi^2\biggl[(11\Nc-2N_f)\left\{\ln\left(\frac{4\pi T}{\varLambda_{\overline {\text{MS}}}}\right)
-\gamma_E\right\} \\
\label{eq:running-coupling}
&~~~+\Nf(4\ln 2-1)-\frac{11\Nc}{2}\biggr]^{-1},\\
\nonumber
m^2_D&= (2\Nc+\Nf)4\pi^2T^2\biggl[(11\Nc-2N_f)\\
\nonumber
&~~~\times\left\{\ln\left(\frac{4\pi T}{\varLambda_{\overline {\text{MS}}}}\right)
-\gamma_E\right\}\\
&~~~+4\Nf \ln 2-\frac{5\Nc^2+\Nf^2+9\Nf/(2\Nc)}{2\Nc+\Nf}\biggr]^{-1},
\end{align}
where $\varLambda_{\overline {\text{MS}}}$ is the renormalization scale in the modified minimal subtraction scheme and $\gamma_E\simeq 0.577$ is Euler's constant.
From Eq.~(\ref{eq:correction-Polyakov}), we see that $l(Q=0)$ exceeds unity due to the renormalization effect.
Now we assume the following relation which reduces to Eq.~(\ref{eq:correction-Polyakov}) at $Q=0$ and $\delta l\ll 1$:
\begin{align}
\label{eq:correction-Polyakov-Q}
l(Q)&= e^{\delta l(Q=0)}\, l^{(0)}(Q).
\end{align}
With $l$ deduced from lattice QCD~\cite{Bazavov:2009zn}, we can determine $l^{(0)}$ from Eq.~(\ref{eq:correction-Polyakov-Q}), and then obtain $Q$ from Eq.~(\ref{eq:l-Q}) using $l^{(0)}$.
These quantities are plotted in Fig.~\ref{fig:Polyakov-Q}, setting $\varLambda_{\overline {\text{MS}}}=T_c/1.35$.
Note that $l^{(0)}$ is still different from unity even around $\sim 3T_c$, where $T_c$ is the pseudo-critical temperature, approximately $155$ MeV in recent lattice computations~\cite{Bazavov:2011nk, Borsanyi:2010cj}.  

We add the following remark concerning the interpolating ansatz, Eq.\,(\ref{eq:correction-Polyakov-Q}):
 If the perturbative correction is not considered (i.e. $\delta l = 0$), then $l(Q)$ gives the upper bound for the Polyakov loop. On the other hand, using 
Eq.\,(\ref{eq:correction-Polyakov-Q}) for not too small $\delta l$ presumably overestimates the perturbative correction and is expected to give a lower bound for $l^{(0)}$.

\subsection{Dynamical (constituent) quark mass}
\label{ssc:quark-mass}
Quark mass generation in QCD implies spontaneous $\chi$SB. The dynamical (constituent) quark mass $m$ is not directly accessible in lattice QCD computations, but it figures as a well-defined quantity in studies of the quark propagator using Dyson-Schwinger equations \cite{Aguilar:2011}
and schematic models of NJL type \cite{Nambu:1961tp}. Starting from massless current quarks, such approaches feature, in the form of a characteristic gap equation, a proportionality between the dynamical quark mass $m$ and the chiral condensate $\langle\bar{q}q\rangle$, realised also at finite temperatures $T$ as indicated in Eq.~(\ref{eq:condensate}). The transition from the 
$\chi$SB (Nambu-Goldstone) phase to chiral symmetry restoration in the Wigner-Weyl phase with $\langle\bar{q}q\rangle = 0$ is second order in the limit of vanishing current quark masses. This transition becomes a continuous but rapid crossover when the explicit chiral symmetry breaking by small non-zero current quark masses is included.

In the present work we make use of the typical behaviour of $m(T)$ derived from NJL {{and PNJL}} models. The input in such models is designed to reproduce vacuum properties of the pion (its decay constant $f_\pi$ and its physical mass $m_\pi$ starting from non-zero current quark masses). In the two-flavour case the emerging dynamical $u$ and $d$ quark masses can be well represented by the following expression:
\begin{align}
m(T)&=\frac{m_0}{2} \left[1-\frac{2}{\pi}\arctan\left(\frac{T-T_0}{\delta}\right)\right]~.
\label{eq:NJLmass}
\end{align}
A standard fit to two-flavour NJL results gives $m_0 \simeq 350$ MeV, $\delta\simeq 20$ MeV and a transition temperature $T_0\simeq 180$ MeV. 
{{The PNJL model features a characteristic entanglement of $m(T)$ reflecting spontaneous (dynamical) $\chi$SB, with the Polyakov loop representing confinement.
While values of $T_0$ found in PNJL calculations are typically larger than those found in NJL models, the corresponding profiles of $m(T)$ are qualitatively similar in all such approaches, apart from shifting the $T$-scale~\cite{Fukushima:2003fw,Islam:2014sea}. 
}}

The $N_f =2$ NJL {{and PNJL}} model calculations yield values of $T_0$ that are {{usually}} higher than the transition temperature determined in recent $N_f = 2 + 1$ lattice QCD computations~\cite{Bazavov:2011nk, Borsanyi:2010cj}.
 Since one of the aims of the present paper is to explore the systematics of quark mass effects on dilepton and photon production rates, we allow variations of $T_0$ between 150 and 180 MeV, and we also examine the influence of different slopes of the chiral crossover controlled by the parameter $\delta$. Four parameter sets will be used, as given in Table \ref{tab:parameter}.
The constituent quark mass $m_0$ at $T=0$ will be kept fixed (at $m_0 = 347 $ MeV) for all
sets A - D. Set B with $T_0=150$ MeV is motivated by the chiral transition temperature $T_c \simeq 155$ MeV found in the lattice QCD computations of Refs. \cite{Bazavov:2011nk, Borsanyi:2010cj}. 
Sets C and D with different values for $\delta$ are used in order to analyze the sensitivity to the steepness of the crossover.
{{With these parameter options the behavior of $m(T)$ in both NJL and PNJL models can be covered,  their main difference being a shift in $T_0$.}}

\begin{figure}[t] 
\begin{center}
\includegraphics[width=0.5\textwidth]{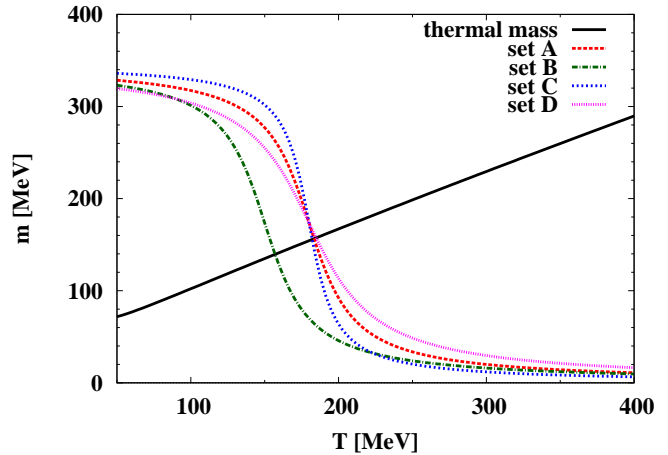}
\caption{The dynamical (constituent) quark mass, Eq. (\ref{eq:NJLmass}) as function of $T$, evaluated using parameter sets A, B, C, and D.
Also plotted for comparison is the thermal quark mass, Eq.~(\ref{eq:thermalmass}). }  
\label{fig:mass-various}
\end{center}
\end{figure}

The temperature dependent quark masses $m(T)$ with these parameter sets are shown in Fig.~\ref{fig:mass-various}.
For later convenience, we also plot the thermal quark mass in the deconfined phase (with 
$\Nc=3$):
\begin{align}
\label{eq:thermalmass}
m_{th}(T) &= \frac{gT}{\sqrt{6}}~,
\end{align}
with $g$ determined by Eq.~(\ref{eq:running-coupling}).
The dynamically generated quark mass tends to exceed the thermal mass at temperature $T\lesssim 200$MeV.

\begin{table}[t]
\caption{Parameters (in MeV) used in the expression (\ref{eq:NJLmass}) for the dynamical quark mass $m(T)$.}
\label{tab:parameter}
\begin{tabular}{l c c c c}
\hline 
  Set & A & B &  C & D  \\ \hline 
$\delta$ & 22 & 22 & 13 & 33\\
$T_0$ &180 &150 & 180 & 180\\
\hline
\end{tabular} 
\end{table}


\section{Dilepton Production}
\label{sec:dilepton}

\begin{figure}[t] 
\begin{center} 
\includegraphics[width=0.15\textwidth]{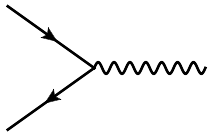}
\caption{Quark-antiquark pair annihilation process contributing to dilepton production.
The solid lines represent the quark or antiquark while the wavy line represents the (timelike) photon that converts into the lepton pair.} 
\label{fig:dilepton-process}
\end{center}
\end{figure}

At leading order of the strong and electromagnetic coupling, the dominant process contributing to the dilepton production rate is the quark-antiquark pair annihilation process followed by pair production of the leptons via the photon.
The amplitude of this process factorizes into a part describing the $q\bar{q}\rightarrow\gamma$ process (shown in Fig.~\ref{fig:dilepton-process}), and a part describing the conversion of the photon into the dilepton~\cite{McLerran:1984ay}. The differential production rate is\footnote{Note that the differential rate $d\varGamma$ is defined so that it is proportional to the phase space element $(d^3p_1/2E_1)\,(d^3p_2/2E_2)$ of the lepton pair, and therefore carries mass dimension four.}
\begin{align}
\label{eq:dilepton-formal}
\frac{d\varGamma}{d^4p}
&=-\frac{\alpha}{24\pi^4p^2}\varPi^{<\mu}{}_{\mu}(p) ,
\end{align} 
where $p_1$ and $p_2$ are the four-momenta of the leptons, $p\equiv p_1+p_2$, and $\alpha\equiv e^2/(4\pi)$.
Here $\varPi^{<\mu}{}_{\nu}(p)$ is the Wightman photon self-energy, which includes the information about the dynamics of the $q\bar{q}\rightarrow\gamma$ process.
The leptons are on-shell ($p^2_1=p^2_2=0$ neglecting the small lepton masses), hence $p$ is time-like ($p^2>0$). 
At leading order we have the following expression for $\varPi^{<\mu}{}_{\mu}(p)$:
\begin{align}
\label{eq:dilepton-expression}
\begin{split}
\varPi^{<\mu}{}_\mu(p)&=
 -\sum_{a, f, {\text{spin}}}\int\frac{d^3\vk_1}{2E_1(2\pi)^3}\int\frac{d^3\vk_2}{2E_2(2\pi)^3}\,\,
 |{\cal M}|^2 \\
&~~~\times (2\pi)^4\delta^{(4)}(p-k_1-k_2)\,n_a(E_1)\,n_{\overline{a}}(E_2) ,
\end{split}
\end{align}
where $f$ is a flavour index running from $1$ to $\Nf$, and $k^0_i=E_{i}\equiv\sqrt{\vk^2_i+m^2(T)}$ are the quark and antiquark energies, $m(T)$ is the temperature-dependent constituent quark mass and $n_{\overline{a}}(E)\equiv [e^{\beta(E+iQ^a)}+1]^{-1}$ is the distribution function for anti-quarks~\cite{Hidaka:2009hs} with a sign change of the imaginary chemical potential as compared to \\$n_a(E)\equiv [e^{\beta(E-iQ^a)}+1]^{-1}$ for quarks. 
The squared matrix element is 
\begin{align}
\begin{split}
\sum_{{\text{spin}}} |{\cal M}|^2
&= 4e^2 q^2_f (p^2+2m^2), 
\end{split}
\end{align}
where we have used $k^2_1=k^2_2=m^2$, and $q_f$ are the electric quark charges in units of $e$ for each flavour $f$.

\subsection{The $|\vp|=0$ case}
\label{ssc:dilepton-p=0}

\begin{figure*}[t] 
\begin{center} 
\includegraphics[width=1.00\textwidth]{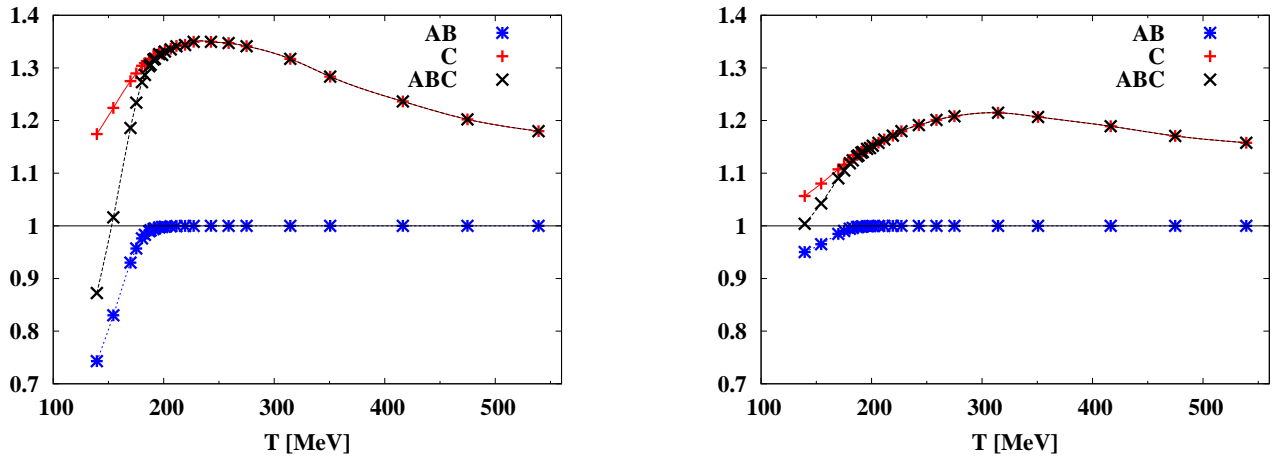}
\caption{The quark mass dependence (displayed by the quantity $AB$) and the Polyakov loop dependence (displayed by $C$) of the dilepton production rate at $|\vp|=0$, as functions of temperature $T$.  Also shown is the $T$ dependence of the product $ABC$.
Parameter set A is used to obtain the input mass $m(T)$. 
In the left panel, $p^0$ is set to 700 MeV while in the right panel, $p^0 = 1$ GeV.
The curves in both panels connecting the computed points are shown to guide the eye.
} 
\label{fig:dilepton-ABC}
\end{center}
\end{figure*}

\begin{figure}[t] 
\begin{center} 
\includegraphics[width=0.50\textwidth]{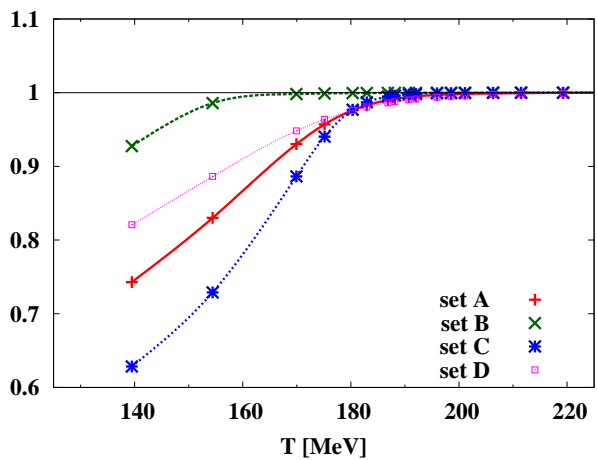}
\caption{Results for the quantity $AB$ using the four parameter sets (A, B, C, and D) of 
Table~\ref{tab:parameter}. The energy $p^0$ is chosen to be 700 MeV.
The curves connecting the computed points are drawn to guide the eye.
} 
\label{fig:dilepton-fitdependence}
\end{center}
\end{figure}

Consider first the case with zero total three-momentum of the lepton pair. 
When $|\vp|=0$ the integration in Eq.~(\ref{eq:dilepton-expression}) can easily be performed.
The leptons are produced back-to-back. The energies of the quark and anti-quark satisfy $E_1=E_2 \equiv E=p^0/2$.
As a result Eqs.~(\ref{eq:dilepton-formal}) and (\ref{eq:dilepton-expression}) yield\footnote{Here the dependence on the number of flavour enters only through the prefactor $\sum_{f}q^2_f$, because we have assumed a common $T$-dependent dynamical quark mass $m$ for all (light) flavours throughout the present work.
For $\Nf=2$ we have $\sum_{f}q^2_f=5/9$.
}
\begin{align}
\label{eq:dilepton-result-p=0}
\begin{split} 
\frac{d\varGamma}{d^4p}&=\frac{\alpha^2}{12\pi^4 E^2} 
 \sum_{f}q^2_f \frac{\sqrt{E^2-m^2}}{E}   \left(E^2+\frac{m^2}{2}\right) \\
&~~~\times \sum_a n_a(E)\,n_{\overline{a}}(E) 
\,{{\theta(E-m)}}.
\end{split}
\end{align}
{{In the case $E<m$ the dilepton production rate vanishes for obvious kinematic reasons.}}
The physical interpretation of this expression is evident.
It {{can conveniently be factorized}} into three parts, coming from the phase space integral, the matrix element squared, and the distribution functions. 
For $\Nc=3$ Eq.~(\ref{eq:dilepton-result-p=0}) becomes
\begin{align}
\label{eq:dilepton-result-p=0-N=3}
\begin{split} 
\frac{d\varGamma}{d^4p}&=\frac{\alpha^2}{4\pi^4} 
 \sum_{f}q^2_f \,A(m) B(m) C(Q)\left({1\over e^{\beta E} +1}\right)^2\\
&~~~\times {{\theta(E-m)}}~, 
\end{split}
\end{align}
where 
\begin{align}
A(m)&\equiv {\sqrt{E^2-m^2(T)}\over E} ~,\nonumber\\
 B(m)&\equiv 1+{m^2(T)\over 2 E^2} ~,\nonumber\\
 C(Q)&\equiv {e^{-2\beta E}+\left[l(Q)+1\right]e^{-\beta E}+1\over e^{-2\beta E}+\left[3l(Q)-1\right]e^{-\beta E}+1}  ~,\nonumber
 \end{align}
 come from these three parts, respectively.
Note that $A$ and $B$ become unity when chiral symmetry is completely restored ($m=0$), while 
$C = 1$ when there is no confinement effect ($l=1$). For $m=0$ our expression reduces to that in one of the author's previous works ~\cite{Gale:2014dfa, DileptonPhoton}.

Only the parts $A$ and $B$ depend explicitly on the quark mass $m(T)$. 
The term $C$ depends on the quark and antiquark energy $E$, but this is in turn fixed by the lepton energy, i.e. $E = p^0/2$.
We also note that the product $AB$ is always less than one as long as $p^0>2m$. This implies that finite $m$ always reduces the dilepton production rate.
By contrast, the background gluon field $Q_a$ enters {{directly}} in part $C$ through the sum over products of quark and antiquark distributions, $n_a(E)\,n_{\overline{a}}(E)$. 
This separation of terms in Eq.~(\ref{eq:dilepton-result-p=0-N=3}) helps making the analysis of $\chi$SB and confinement effects transparent\footnote{{Given the entanglement of confinement and $\chi$SB as it is realized in the PNJL model, $m$ and $Q$ are not independent quantities.
It is nonetheless instructive to analyze the effects of confinement and $\chi$SB on the dilepton production separately, and so it is convenient in the present context to treat $m$ and $Q$ as independent quantities.
}}, 
{{especially in comparison with previous results of a semi-QGP model calculation~\cite{Gale:2014dfa, DileptonPhoton} in which the $\chi$SB effect was not considered.
The difference is then primarily in the function $AB$.}}

The quantities $AB$ and $C$ are shown as functions of $T$ in the left panel of Fig.~\ref{fig:dilepton-ABC}. In the given example $p^0$ is fixed at 700 MeV, where we have used parameter set A for the evaluation of the quark mass $m$.
The $AB$ plot displays the $m$ dependence while the latter plot contains information about the dependence on $l$.
From the $AB$ plot one observes that this quantity is significantly smaller than unity in the region $T<180$ MeV: the constituent quark mass suppresses the dilepton production rate by modifying the available phase space and the matrix element.
When $T>180$ MeV $AB$ quickly approaches unity because around this temperature, $m$ decreases rapidly as shown in Fig.~\ref{fig:mass-various} and becomes negligible compared to the energy $E=p^0/2=350$ MeV appearing in $AB$.
From the plot of $C$ one observes that the effect of the Polyakov loop does not suppress the dilepton production, but even slightly enhances it.
This unexpected behaviour has already been discussed previously in Refs.~\cite{DileptonPhoton, Gale:2014dfa}, referring to the fact that the initial state (the quark-antiquark pair) is color-singlet.

The product $ABC$ indicates the modification of the dilepton production rate compared to the case with $l=1$ and $m=0$. It shows the combined effect of $m$ and the Polyakov loop. 
Since $C > 1$, the suppression coming from finite $m$ in the region $T<180$ MeV is balanced by the enhancement due to the Polyakov loop. 
For $T >$ 180 MeV the factor $AB$ approaches unity and $ABC$ is nearly equal to $C$.
With increasing delepton energy $p^0$ the influence of $m$ becomes less significant. This is illustrated by comparing the left and right panels of Fig.~\ref{fig:dilepton-ABC}. At $p^0=1$ GeV,  
$AB$ differs only slightly from unity even when $T<180$ MeV.

We also examine the dependence of the dilepton production rate on the four different
parameter sets summarized in Table~\ref{tab:parameter}. The result for $AB$ is shown in Fig.~\ref{fig:dilepton-fitdependence} using $p^0=700$ MeV.
Lowering $T_c$ increases the dilepton rate as the comparison between ``set B" and ``set A" results shows. This trend is expected since, as $T_c$ is lowered, $m$ becomes smaller in the relevant range and so the suppression effect on dilepton production is reduced.
On the other hand, 
if the chiral crossover transition proceeds with a steeper slope around $T_c$, the quark mass $m$ below the transition temperature tends to be larger and the suppression effect induced by $m$ is enhanced, as can be seen from the comparison of the plots for ``set A", ``set C", and ``set D". 
Above the transition temperature the quark mass quickly decreases and the effect is not visible in the figure.

\subsection{The $|\vp|\neq 0$ case}

For non-vanishing three-momentum of the lepton pair, $|\vp|\neq 0$, using 
\begin{align}
\delta(p^0-E_1-E_2)
&= \frac{|p^0-E_1|}{|\vp||\vk_1|}\delta(\cos\theta-\cos\theta_0),
\end{align} 
where $\cos\theta_0\equiv -(p^2-2p^0 E_1)/(2|\vp||\vk_1|)$ and $\theta$ is the angle between $\vp$ and $\vk_1$,
Eqs.~(\ref{eq:dilepton-formal}) and (\ref{eq:dilepton-expression}) give:
\begin{align}
\label{eq:dilepton-result-finitep}
\begin{split}
\frac{d\varGamma}{d^4p}
&=\frac{\alpha^2}{12\pi^4p^2} 
\sum_{a, f} q^2_f \, {p^2+2m^2\over |\vp|}\, {{\theta(p^2-4m^2)}}   \\
&~~~\times \int^{E_+-p^0/2}_{E_--p^0/2} dE\, 
n_a\left(\frac{p^0}{2}+E\right) n_{\overline{a}}\left(\frac{p^0}{2}-E\right) ,
\end{split}
\end{align}
where 
\begin{align}\
E_{\pm}\equiv& \sqrt{k^2_{\pm}+m^2}~, \nonumber \\
k_\pm\equiv& {1\over 2}\left(|\vp|\pm p^0\sqrt{1-{4m^2\over p^2}}\right)~. \nonumber
\end{align}
{{Kinematics implies that the dilepton production rate vanishes for $p^2<4m^2$.}}
Performing the integration one arrives at: 
\begin{align}
\begin{split} 
\frac{d\varGamma}{d^4p}
&=\frac{\alpha^2}{12\pi^4p^2} 
\sum_{a, f} q^2_f  (p^2+2m^2)\frac{T}{|\vp|} \nb(p^0)\,{{\theta(p^2-4m^2)}}   \\
&~~~\times  \Biggl[ \ln\left(\frac{1+e^{-\beta(E_+-iQ_a)}}{1+e^{-\beta(E_- -iQ_a)}}\right) \\
&~~~ -\ln\left(\frac{e^{-\beta p^0}+e^{-\beta(E_+-iQ_a)}}{e^{-\beta p^0}+e^{-\beta(E_- -iQ_a)}}\right)
 \Biggr] .
\end{split}
\end{align}
In contrast to the $|\vp|=0$ case, the effects of the Polyakov loop and of the quark mass $m$ are not well separated any more.
In the limit $m=0$ we reproduce the corresponding expression in Ref.~\cite{DileptonPhoton, Gale:2014dfa}.

In the $\Nc=3$ case, we have 
\begin{align}
\begin{split} 
\frac{d\varGamma}{d^4p}
&=\frac{\alpha^2}{12\pi^4} 
\sum_{f} q^2_f  \, \nb(p^0)\, {{\theta(p^2-4m^2)}}\\
&~~~\times 3\left[1+ 2\frac{T}{|\vp|}\ln\frac{1+e^{-\beta(p^0+|\vp|)/2}}{1+e^{-\beta(p^0-|\vp|)/2}}\right]
\tilde{B}(m) f_{l\overline{l}}(Q,m)  ,
\end{split}
\end{align} 
where $\tilde{B}(m)\equiv (p^2+2m^2)/p^2$ and $f_{l\overline{l}}(Q,m)\equiv \tilde{f}_{l\overline{l}}(Q,m)/\tilde{f}_{l\overline{l}}(0,0)$, with
\begin{widetext}
\begin{align}
\begin{split}
\tilde{f}_{l\overline{l}}(Q,m)& \equiv 
\ln\left(\frac{1+3l\, e^{-\beta E_+}+3l\, e^{-2\beta E_+}+e^{-3\beta E_+}}
{1+3l\, e^{-\beta E_-}+3l\, e^{-2\beta E_-}+e^{-3\beta E_-}}\right) 
 -   \ln\left(\frac{e^{-3\beta p^0}+3l \,e^{-2\beta p^0}e^{-\beta E_+}+3l \,e^{-\beta p^0}e^{-2\beta E_+}+e^{-3\beta E_+}}
{e^{-3\beta p^0}+3l \,e^{-2\beta p^0}e^{-\beta E_-}+3l \,e^{-\beta p^0}e^{-2\beta E_-}+e^{-3\beta E_-}}\right).
\end{split}
\end{align}
\end{widetext}
Note that $\tilde{B} = 1$ when $m=0$, and $f_{l\overline{l}}(Q,m)$ becomes unity when $l=1$ and $m=0$.

\begin{figure*}[t] 
\begin{center} 
\includegraphics[width=1.00\textwidth]{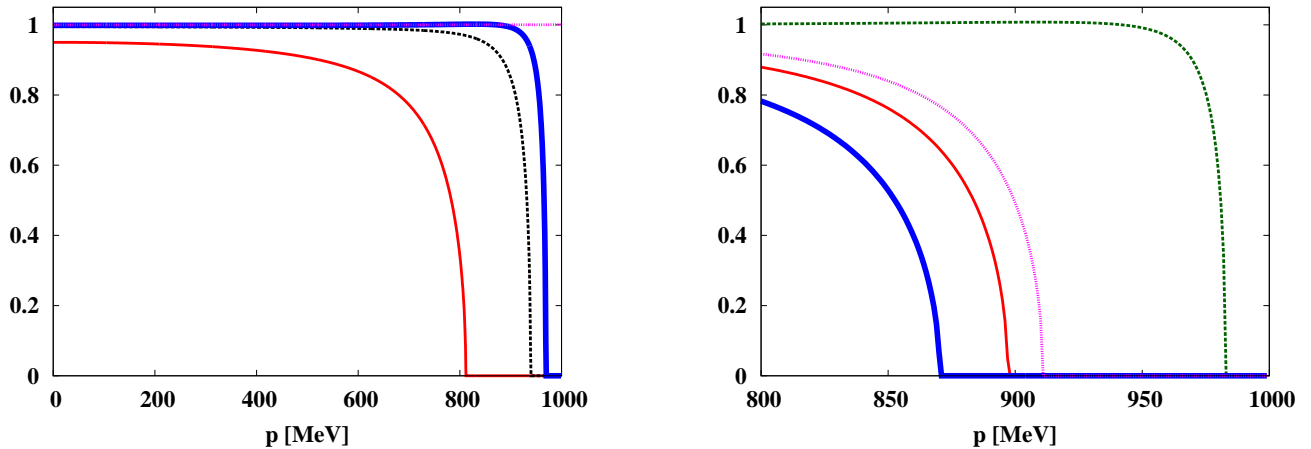}
\caption{Ratio of dilepton production rate, ${d\varGamma}/{d^4p}/({d\varGamma}/{d^4p}|_{m=0})$, with and without inclusion of dynamical quark mass $m(T)$, as a function of $|\vp|$ with $p^0 =$1 GeV.
Left panel: using parameter set A.
Results are plotted at different temperatures. $T=$ 140 MeV: solid (red) curve; $T=$ 180 MeV: dashed (black) curve; $T=$ 190 MeV: thick solid (blue) curve; and $T=$ 540 MeV: dotted (magenta) curve. 
Right panel: same quantity at fixed temperature $T =$ 170 MeV for different parameter sets.
Parameter set A: solid (red) curve; set B:  dashed (green) curve; set C: thick solid (blue) curve; set D: dotted (magenta) curve.
} 
\label{fig:dilepton-finitep}
\end{center}
\end{figure*}

The effect of the Polyakov loop $l$ is analyzed in Refs.~\cite{Gale:2014dfa, DileptonPhoton}.
Here we focus on the role of the dynamical quark mass, $m(T)$.
The quantity $\tilde{B}(m) f_{l\overline{l}}(Q,m)/ f_{l\overline{l}}(Q,0)$, which is the ratio of the dilepton production rate at finite $m$ to that at $m=0$ (${d\varGamma}/{d^4p}/({d\varGamma}/{d^4p}|_{m=0})$), is shown in Fig.~\ref{fig:dilepton-finitep} as a function of $|\vp|$ at three temperatures around $T_0$ (140, 180, 190 MeV), and at one temperature ($T = 540$ MeV) that is significantly larger than $T_0$. In the left panel,
$p^0$ is fixed at 1GeV and the parameter set A is used. Suppression effects on dilepton production due to non-zero quark masses are marginal at low momentum $|\vp|$ but become significant at higher momenta. Complete suppression occurs at the limit, $p^2 = 4m^2$, due to the kinematic constraint.
 At lower temperature the quark mass $m(T)$ is larger than at higher temperature, hence the suppression is more significant.

The parameter dependence of the dilepton production rate encoded in $\tilde{B}(m) f_{l\overline{l}}(Q,m)/ f_{l\overline{l}}(Q,0)$ is plotted in the right panel of Fig.~\ref{fig:dilepton-finitep} as a function of $|\vp|$, at $T=$ 170 MeV and $p^0=$1 GeV, using the parameter sets A, B, C, and D. The differences in $m(T)$ translate directly to the momentum ($p_k$) at which $\tilde{B}(m) f_{l\overline{l}}(Q,m)/ f_{l\overline{l}}(Q,0)$ vanishes because of the kinematic constraint:
$p_k|_C<p_k|_A<p_k|_D$, and $p_k|_B$ is larger than $p_k|_A$.
Since $p_k=\sqrt{(p^0)^2-4m^2}$ is determined entirely by $m(T)$, this tendency is understood in the same way as explained in the discussion of the quantity $AB$ in the previous subsection.

\section{Photon Production}
\label{sec:photon}  

\begin{figure}[t] 
\begin{center} 
\includegraphics[width=0.35\textwidth]{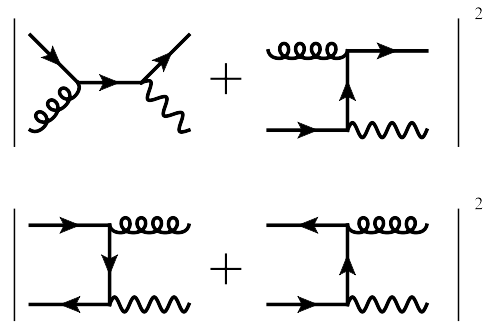} 
\caption{Basic photon production processes: Compton scattering and pair annihilation.  
The curly line represents the gluon.
} 
\label{fig:photon-2to2}
\end{center}
\end{figure}

At leading order the dominant contributions to the photon production rate come from Compton scattering $(qg \rightarrow q\gamma)$ and pair annihilation $(q\bar{q}\rightarrow g\gamma)$ \cite{Kapusta:1991qp, Staadt:1985uc, Baier:1991em, DileptonPhoton, Gale:2014dfa}. The Landau-Pomeranchuk-Migdal (LPM) effect~\cite{Baym:2006qf, Arnold:2001ba, Aurenche:2000gf} also contributes at the same order when $m=0$. 
In this paper we neglect the contribution from the LPM mechanism. This 
is justified when $m$ is large compared to the thermal quark mass $(m \gg m_{th} \sim gT)$.  It is also justified, even when $m=0$, in the  large $\Nc$ limit~\cite{DileptonPhoton}.
The two basic processes, quark-gluon
Compton scattering and pair annihilation of quarks, are illustrated in Fig.~\ref{fig:photon-2to2}.
The amplitudes of these processes are infrared-singular in the limit of small energy and momentum of the exchanged quark. These singularities are naturally regularized 
when the constituent quark mass is finite. The resulting photon production rate has a characteristic logarithmic dependence on $m$ as will be shown later.
Even when $m=0$ the thermal quark mass acts to regularize the amplitudes such that they depend logarithmically on $m_{th}(T)$. Which mass acts prominently as a regulator depends on which one is larger at a given temperature.
With the purpose of this paper to investigate finite-mass effects systematically, we calculate the photon production rates with finite dynamical quark masses and compare versus results with $m=0$ but using a resummed quark propagator, which contains the information on the thermal quark mass.

\subsection{Compton scattering}
Consider first the contribution from the Compton scattering process.
Within the kinetic theory, the photon production rate from this mechanism is expressed as
\begin{align}
\label{eq:Compton-expression}
\begin{split}
\left. E\frac{d\varGamma}{d^3\vp}\right|_{\text{Comp}}&= \frac{1}{2(2\pi)^3} \sum
\int\frac{d^3\vk}{(2\pi)^3}\int\frac{d^3\vk'}{(2\pi)^3}\int\frac{d^3\vp'}{(2\pi)^3} \\
&~~~\times \frac{|M|^2}{8E_\vk |\vk'|E_{\vp'}}
(2\pi)^4\delta^{(4)}(k+k'-p-p')\\
&~~~\times\nf{}_a(E_\vk)\,\nb{}_{cd}(|\vk'|)\,
 [1-\nf{}_{b}(E_{\vp'})]~,
\end{split}
\end{align}
where the sum is over quark flavours $f$, the spins of all the particles, and color indices ($a, b, c, d$).
$M$ is the matrix element of the Compton scattering process, the explicit expression of which will be given later. We consider a real photon with $p^0=|\vp|\equiv E$
and focus on the $E\gg T$ case relevant for high-energy heavy-ion collisions.
The Boltzmann approximation will be used in the initial state:
in Eq.~(\ref{eq:Compton-expression}), we replace $\nf(E_\vk-iQ_a)\,\nb(|\vk'|-iQ_c+iQ_d)$ by $\exp[-\beta(E_\vk+|\vk'|-iQ_a-iQ_c+iQ_d)]$.
This is justified when $E_\vk$, $|\vk'|\gg T$. It is known that this treatment gives the correct result at leading-log order~\cite{Kapusta:1991qp, Baier:1991em}. 
With this approximation Eq.~(\ref{eq:Compton-expression}) leads to the following expression: 
\begin{align} 
\begin{split}
\left. E\frac{d\varGamma}{d^3\vp}\right|_{\text{Comp}}
&=  \frac{1}{32(2\pi)^6}\frac{T}{E} e^{-\beta E}\sum_{} 
\int^\infty_{0} d\tilde{s} \int^{\tilde{t}_+}_{\tilde{t}_-} d\tilde{t}  \frac{|M|^2}{\tilde{s}} \\
&~~~\times e^{i\beta(Q_a-Q_b+Q_c-Q_d)} \\
&~~~\times  \ln\left(1+e^{-\beta(m^2E/\tilde{s}+\tilde{s}/(4E)-iQ_b)}\right)~. 
\end{split}
\end{align}
Here we have introduced the Mandelstam variables: $s\equiv (k+k')^2$ and $t\equiv (p-k)^2$.
The variables with tilde are defined as $\tilde{s}\equiv s-m^2$ and $\tilde{t}\equiv t-m^2$,
and we have introduced $\tilde{t}_+\equiv -m^2\tilde{s}/s$ and $\tilde{t}_-\equiv -\tilde{s}$.
For details concerning this change of integration variables, see Refs.~\cite{Staadt:1985uc, DileptonPhoton}.

The squared matrix element after spin summations is given by:
\begin{align}
\begin{split} 
\sum_{\text{spin}} |M|^2 
&= 32\,e^2g^2q^2_f \,[(t^{cd})_{ba}]^2 
\Biggl[\frac{1}{\tilde{s}^2}\left(-\frac{\tilde{s}\tilde{t}}{2}+m^2\tilde{s}+2m^4\right) \\
&+\frac{1}{\tilde{t}^2}\left(-\frac{\tilde{s}\tilde{t}}{2}+m^2\tilde{t}+2m^4\right) 
+\frac{m^2}{\tilde{s}\tilde{t}}\left(\tilde{s}+\tilde{t}+4m^2\right)
\Biggr],
\end{split}
\end{align}
where $(t^{cd})_{ba}\equiv {\cal P}^{cd}_{ba}/\sqrt{2}$ with ${\cal P}^{cd}_{ba} \equiv \delta^{c}_{b}\delta^{d}_{a}-\delta^{cd}\delta_{ba}/\Nc$ encodes the color structure of the quark-gluon vertex in  double-line notation.
The first term is the contribution from the $s$-channel process, the second from $t$-channel exchange, and the third term comes from the interference of these two amplitudes, respectively.
One finds:
\begin{align}
\label{eq:photon-compton} 
\begin{split} 
\left. E\frac{d\varGamma}{d^3\vp}\right|_{\text{Comp}}
&= \frac{\alpha\alpha_s}{8\pi^4} (\Nc^2-1) \frac{T}{E}\, e^{-\beta E}\sum_fq^2_f \\
&~~~\times \int^\infty_{0} d\tilde{s} \, G(\tilde{s},m^2)  \\
&~~~\times \sum^\infty_{n=1} {(-1)^{n+1}\over n}\,l_n\,e^{-n\beta(m^2E/\tilde{s}+\tilde{s}/(4E))}~, 
\end{split} 
\end{align}
where $\alpha_s\equiv g^2/(4\pi)$ and
\begin{align}
\begin{split}
G(\tilde{s},m^2)&\equiv \frac{\tilde{s}+2m^2}{4(\tilde{s}+m^2)^2}\tilde{s} +4\frac{m^2}{\tilde{s}} \\
&~~~-\left(-\frac{1}{2}+2\frac{m^2}{\tilde{s}}+\frac{4m^4}{\tilde{s}^2}\right)\ln\left|1+\frac{\tilde{s}}{m^2}\right|.
\end{split}
\end{align}
In the deconfined phase with $l_n=1$, this expression becomes:
\begin{align}
\label{eq:photon-compton-deconfine} 
\begin{split} 
\left. E\frac{d\varGamma}{d^3\vp}\right|_{\text{Comp}}
&= \frac{\alpha\alpha_s}{8\pi^4} (\Nc^2-1) \,\frac{T}{E} e^{-\beta E}\sum_fq^2_f \int^\infty_{0} d\tilde{s}   \\
&~~~\times G(\tilde{s},m^2)\ln\left(1+e^{-\beta(m^2E/\tilde{s}+\tilde{s}/(4E))}\right) ~.
\end{split} 
\end{align}
In the limit $m^2\rightarrow 0$ the logarithmic term in $G(\tilde{s},m^2)$ dominates over the other pieces:
\begin{align} 
\label{eq:photon-compton-LL}
\begin{split} 
\left. E\frac{d\varGamma}{d^3\vp}\right|_{\text{Comp}}
&\simeq \frac{\alpha\alpha_s}{16\pi^4}(\Nc^2-1) \frac{T}{E} e^{-\beta E}\sum_fq^2_f 
\int^\infty_{0} d\tilde{s}   \\
&~~~\times \sum^\infty_{n=1} \frac{(-1)^{n+1}}{n} e^{-n\beta \tilde{s}/(4E)}\,l_n
\ln\left|1+\frac{\tilde{s}}{m^2}\right| \\
&\simeq \frac{\alpha\alpha_s}{4\pi^4} (\Nc^2-1) \,T^2 e^{-\beta E}\sum_fq^2_f  \\
&~~~\times \sum^\infty_{n=1} \frac{(-1)^{n+1}}{n^2} \,l_n
\ln\frac{ET}{m^2} ~.
\end{split}
\end{align}
This expression agrees with the one in leading-log order when regarding $m^2$ as the infrared cutoff $\mu^2$ used in Ref.~\cite{DileptonPhoton}.
It also suggests that $m$ works as an infrared regulator, as was mentioned. 

Consistency with neglecting the LPM contribution requires to take the large $\Nc$ limit.
The production rate is then given by Eq.~(\ref{eq:photon-compton}) with the replacement $\Nc^2-1\rightarrow \Nc^2$ in the prefactor.
By extrapolating this expression to the $\Nc=3$ case, and inserting $l_n$ from Eq.~(\ref{eq:ln-Q}),
we have
\begin{align} 
\label{eq:Compton-result}
\begin{split} 
\left. E\frac{d\varGamma}{d^3\vp}\right|_{\text{Comp}}
&= \frac{\alpha\alpha_s}{4\pi^4} T^2 e^{-\beta E}\sum_fq^2_f \,X,
\end{split}
\end{align}
with the dimensionless function
\begin{align}
\begin{split}
X(T) &\equiv \frac{3}{2ET}  
 \sum^\infty_{n=1} {(-1)^{n+1}\over n}[1+2\cos( \beta n Q)] \\
 & ~~~\times\int^\infty_{0} d\tilde{s} \,G(\tilde{s},m^2)\,e^{-n\beta(m^2E/\tilde{s}+\tilde{s}/(4E))} . 
\end{split}
\end{align}

\subsection{Pair annihilation}

In a way similar to the case of the Compton scattering process, the photon production rate due to pair annihilation can be written as:
\begin{align}
\label{eq:pair-expression}
\begin{split}
\left.E\frac{d\varGamma}{d^3\vp}\right|_{\text{pair}}&= \frac{1}{2(2\pi)^3} \sum_{} 
\int\frac{d^3\vk}{(2\pi)^3}\int\frac{d^3\vk'}{(2\pi)^3}\int\frac{d^3\vp'}{(2\pi)^3} \\
&~~~\times\frac{|M|^2}{8E_\vk E_{\vk'}|\vp'|}
(2\pi)^4\delta^{(4)}(k+k'-p-p')\\
&~~~\times\nf{}_{a}(E_\vk)\,\nf{}_{\overline{b}}(E_{\vk'})\, [1+\nb{}_{cd}(|\vp'|)]\\
&\simeq \frac{-1}{32 (2\pi)^6}\frac{T}{E}\,e^{-\beta E}\sum_{} \int^\infty_{4m^2} ds \int^{\tilde{t}_+}_{\tilde{t}_-} d\tilde{t} 
   \,\frac{|M|^2}{s} \\
   &~~~\times e^{i\beta(Q_a-Q_b-Q_c+Q_d)} \\ 
   &~~~\times \ln\left[1-e^{-\beta s/(4E)}e^{i\beta(Q_c-Q_d)}\right], 
\end{split}
\end{align}
where we have used again the Boltzmann approximation in the initial state and introduced $u\equiv (p-k')^2$, $\tilde{u}\equiv u-m^2$, and $\tilde{t}_\pm\equiv -s[1\mp\sqrt{1-4m^2/s}]/2$. 
The squared matrix element summed over spins is:
\begin{align} 
\begin{split} 
\sum_{\text{spin}}|M|^2
&=8e^2g^2q^2_f [(t^{dc})_{ba}]^2\Biggl[
\frac{1}{\tilde{t}^2}\left(\tilde{t}\tilde{u}-2m^2\tilde{t}-4m^4\right) \\
&+\frac{1}{\tilde{u}^2}\left(\tilde{u}\tilde{t}-2m^2\tilde{u}-4m^4\right)
+\frac{2m^2}{\tilde{t}\tilde{u}}\left(s-4m^4\right)
\Biggr]. 
\end{split}
\end{align}
Using $s+\tilde{t}+\tilde{u}=0$, the result reads
\begin{align}
\label{eq:photon-pair-result}
\begin{split} 
\left.E\frac{d\varGamma}{d^3\vp}\right|_{\text{pair}}
&= \frac{\alpha\alpha_s}{16\pi^4}\frac{T}{E}e^{-\beta E}\sum_{f}q^2_f 
 \int^\infty_{4m^2} ds  \\
&\times\sum^\infty_{n=1}\frac{1}{n}\,e^{-sn\beta/(4E)}
 (\Nc^2\,l_n^2 -1)\,H(s,m^2),
\end{split}
\end{align}
where
\begin{align}
\begin{split}
H(s,m^2) &\equiv \left(1+\frac{4m^2}{s}\right)\sqrt{1-\frac{4m^2}{s}} \\
&~~~+\left(1+\frac{4m^2}{s}-\frac{8m^4}{s^2}\right)\\
&~~~\times\ln\left|\frac{1+\sqrt{1-4m^2/s}}{1-\sqrt{1-4m^2/s}}\right|.
\end{split}
\end{align}
In the deconfined phase with $l_n = 1$, Eq. (\ref{eq:photon-pair-result}) reduces to
\begin{align}
\label{eq:photon-pair-result-deconfine}
\begin{split} 
\left.E\frac{d\varGamma}{d^3\vp}\right|_{\text{pair}}
&= \frac{\alpha\alpha_s}{16\pi^4}\,\frac{T}{E}e^{-\beta E}\sum_{f}q^2_f \,(\Nc^2-1) \\
&~~~\times \int^\infty_{4m^2} ds 
\left| \ln\left(1-e^{-s\beta/(4E)}\right)\right| \,H(s,m^2).
\end{split}
\end{align}
In the limit $m^2\rightarrow 0$, we have
\begin{align}
\label{eq:photon-pair-LL}
\begin{split} 
\left. E\frac{d\varGamma}{d^3\vp}\right|_{\text{pair}}
&\simeq \frac{\alpha\alpha_s}{4\pi^4} T^2 \,e^{-\beta E}\sum_fq^2_f    
 \sum^\infty_{n=1} \frac{1}{n^2} (\Nc^2\,l^2_n-1)\\
&~~~\times\ln\frac{ET}{m^2} .
\end{split}
\end{align}
which again agrees with the leading-log result~\cite{DileptonPhoton}.

In the large $\Nc$ limit, Eq.~(\ref{eq:photon-pair-result}) becomes
\begin{align}
\begin{split} 
\left.E\frac{d\varGamma}{d^3\vp}\right|_{\text{pair}}
&= \frac{\alpha\alpha_s}{16\pi^4}\,\frac{T}{E}e^{-\beta E}\sum_{f}q^2_f 
 \int^\infty_{4m^2} ds \\
&\times \sum^\infty_{n=1}\frac{1}{n} \,e^{-sn\beta/(4E)} \Nc^2\,l_n^2  \,H(s,m^2)~.
\end{split}
\end{align}
Inserting $\Nc=3$, this expression becomes
\begin{align}
\label{eq:photon-pair-result-N=3}
\begin{split}  
\left.E\frac{d\varGamma}{d^3\vp}\right|_{\text{pair}}
&= \frac{\alpha\alpha_s}{4\pi^4}\,T^2 e^{-\beta E}\sum_{f}q^2_f\,
Y,
\end{split}
\end{align}
with the dimensionless function  
\begin{align}
\begin{split}
Y(T) &\equiv \frac{1}{4ET} \int^\infty_{4m^2} ds 
\sum^\infty_{n=1}\frac{1}{n}e^{-sn\beta/(4E)} \\
&~~~\times \left[2\cos(n\beta Q)+1\right]^2 \,H(s,m^2).
\end{split}
\end{align}

\subsection{Numerical results}
\label{ssc:photon-result}

\begin{figure}[t] 
\begin{center} 
\includegraphics[width=0.50\textwidth]{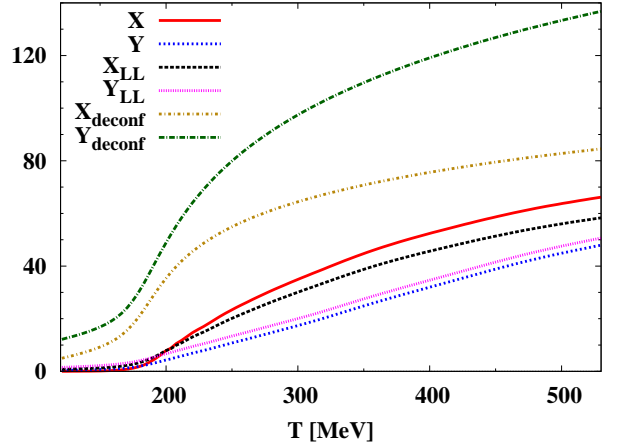} 
\caption{The quantities $X$ and $Y$ as functions of $T$ with $E=2$ GeV.
The parameter set A is used.
Shown for comparison are results obtained in the leading-log approximation (LL) and in the completely deconfined phase ($ X_{\rm deconf}, Y_{\rm deconf}$).
} 
\label{fig:photon-XY}
\end{center}
\end{figure}

\begin{figure}[t] 
\begin{center}
\includegraphics[width=0.50\textwidth]{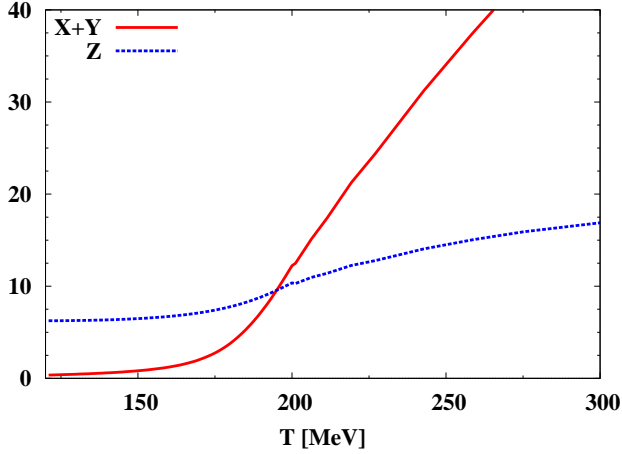} 
\caption{The functions $X+Y$ and $Z$ at various temperatures $T$ with $E=2$ GeV.
The parameter set A is used.
The solid (red) line represents $X+Y$ while the dotted (blue) line represents $Z$. 
} 
\label{fig:photon-XY-Z}
\end{center}
\end{figure}


We proceed by first analyzing separately the contributions from Compton scattering and pair annihilation.
In Eqs.~(\ref{eq:Compton-result}) and (\ref{eq:photon-pair-result-N=3}), the prefactors $\alpha\alpha_s T^2 e^{-\beta E}\sum_f q^2_f/(4\pi^4)$ are common, so we plot the $T$-dependent functions $X$ and $Y$ in Fig.~\ref{fig:photon-XY}.
The energy is set to $E = 2$ GeV, and the parameter set A is used to generate $m(T)$.
As Polyakov loop input 
a fitting function for the lattice data of $l(T)$, plotted in Fig.~\ref{fig:Polyakov-Q}, has been used. Comparing these quantities with the corresponding ones at the leading-log order (Eqs.~(\ref{eq:photon-compton-LL}) and (\ref{eq:photon-pair-LL}) at large $\Nc$ limit), we see that the $T$ dependence can be well described by the leading-log approximation when $T$ is larger than about 250 MeV. At such temperatures $m$ becomes negligible which makes the leading-log approximation more accurate.
Also shown in the same figure
are $X$ and $Y$ in the deconfined limit with $l \equiv 1$ (Eqs.~(\ref{eq:photon-compton-deconfine}) and (\ref{eq:photon-pair-result-deconfine}) at large $\Nc$ limit). 
One observes that the Polyakov loop effect suppresses these quantities significantly at all $T$.
Note that the contribution from pair annihilation is larger than that from the Compton scattering process in the limit $l \equiv 1$, whereas this is not the case when incorporating the Polyakov loop. This behaviour derives from the fact that $X$ is linear in $l_n$ while $Y$ involves $(l_n)^2$, which makes the suppression of $Y$ due to the Polyakov loop more significant.

In order to examine the effect of the dynamical quark mass we compare results with finite $m$ to those obtained using $m=0$.
With inclusion of $m(T)$ the total photon production rate is given by 
\begin{align} 
\label{eq:photon-result}
\left.E\frac{d\varGamma}{d^3\vp}\right|
&= \frac{\alpha\alpha_s}{4\pi^4}\,T^2 \,e^{-\beta E}\sum_{f}q^2_f\,
(X+Y).
\end{align}
In the $m=0$ case the infrared singularity is regularized by the thermal quark mass. The calculation requires a resummation, the derivation of which
is given in the appendix. Here we just quote the result, 
adding Eqs.~(\ref{eq:photon-compton-largeN-N=3}), (\ref{eq:photon-pair-largeN-N=3}), and (\ref{eq:photon-soft-largeN-N=3}):
\begin{align}
\label{eq:photon-result-m=0} 
\begin{split} 
\left. E\frac{d\varGamma}{d^3\vp}\right|_{\text{massless}}
&= \frac{\alpha\alpha_s}{4\pi^4}\,T^2 \,e^{-\beta E}\sum_fq^2_f\, 
Z,
\end{split}
\end{align}
where 
\begin{align}
\label{eq:Z-result-largeN}
\begin{split} 
Z(T)&\equiv 
\pi^2\left(6 q^2-8 q+{9\over 4}\right)
\left(\ln\left(\frac{4E}{g^2T}\right) -\gamma_E-1.32516\right)
\\
&-\pi^2\left(15 q^2-8 q+{9\over 8}\right)
-2.5087\\
&-2\sum^\infty_{n=1} \frac{\ln n}{n^2}
\Bigl[\left(3(-1)^{n+1}+2\right)\cos(2\pi nq)+\cos(4\pi nq)\Bigr]  \\
&-\pi^2
\Biggl[\left({3\over 2}-6 q+6 q^2\right)\ln \left(\frac{3}{16}-{3\over 4} q+{3\over 4} q^2\right)\\
&~~~~~~~~~~~~~~~~+\left(\frac{3}{4}-2q\right)\ln \left(\frac{3}{16}-\frac{q}{2}\right)
\Biggr],
\end{split}
\end{align}
which is found using the expression for large $\Nc$ and inserting $\Nc=3$.
Here $q = Q/(2\pi T)$.
Note that although an arbitrary cutoff $\mu$ is introduced in the derivation, the result is
independent of $\mu$.

The $T$-dependent functions $X+Y$ and $Z$ are shown in Fig.~\ref{fig:photon-XY-Z}.
One observes that $X+Y$ is significantly smaller (larger) than $Z$ when the temperature is smaller (larger) than about 190 MeV, close to the transition temperature $T_0$ of parameter set A. 
This behavior of $X+Y$ can be understood as follows.
In the low temperature region, $m$ is of order or larger than $T$.
From the exponential part of Eq.~(\ref{eq:photon-compton}) it follows that the dominant contribution to $X$ comes from the region $\tilde{s}\sim Em$ so that
the suppression factor $e^{-m/T}$ results in $X$.
As for the pair annihilation contribution, the lower bound of the integral in Eq.~(\ref{eq:photon-pair-result}) gives the exponential suppression factor $e^{-m^2/(ET)}$ in $Y$ when $m$ is large.
On the other hand, when $T$ exceeds about 190 MeV, $m$ becomes small and
makes the leading-log approximation (Eqs.~(\ref{eq:photon-compton-LL}) and (\ref{eq:photon-pair-LL})) more accurate. This leads to an enhancement of $X+Y$ compared to $Z$ because of the factor $\ln(ET/m^2)$.

\begin{figure}[t] 
\begin{center}
\includegraphics[width=0.50\textwidth]{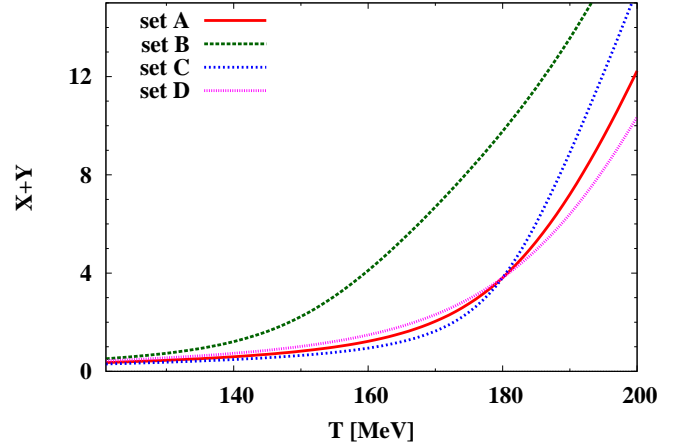} 
\caption{The sum $X+Y$ calculated using the parameter sets $A, B, C, D$ as a function of $T$ at $E=2$ GeV.
} 
\label{fig:photon-parameterdependence}
\end{center}
\end{figure}

While these considerations remain valid, the detailed high-$T$ behavior needs nonetheless to be examined with greater care.
When $m$ is much smaller than the thermal quark mass, which is realized for $T\geq 200$ MeV according to Fig.~\ref{fig:mass-various}, this thermal mass works as an infrared regulator instead of $m$, and thus the role of the leading factor $\ln(ET/m^2)$ appearing in $X+Y$ is now to be replaced by $\ln(E/(g^2T))$ in $Z$. At temperatures well above 200 MeV it is therefore appropriate to use $Z$ rather than $X+Y$. In summary, for practical applications we suggest to use $X+Y$ for  $T<T_m$ and $Z$ for $T>T_m$, where $T_m$ is the temperature at which $X+Y$ equals $Z$.

Next we examine how variations of the temperature $T_0$ and of the steepness $\delta$ of the chiral crossover transition influence the photon production rate.
Fig.~\ref{fig:photon-parameterdependence} shows the quantity $X+Y$, using the parameter sets $A$, $B$, $C$, and $D$, as functions of $T$. 
We focus on the low $T$ region ($T<T_m$) where the results including the dynamical quark mass $m$ are expected to be reliable.
Varying $\delta$ changes the photon production rate only moderately, as can be seen by looking at $X+Y$ using the parameter sets $A$, $C$, and $D$ with $T_0 = 180$ MeV.
By contrast, varying $T_0$ induces a more significant change in $X+Y$ as can be seen by using the parameter set  $B$ with $T_0 = 150$ MeV. The shift of the quark mass profile $m(T)$ to a lower transition temperature implies an increase of $X+Y$ already around $T\sim T_0$ where the quark mass starts to drop rapidly.

\section{Summary and Concluding Remarks}
\label{sec:summary}

In this work dilepton and photon production rates in the quark-gluon phase close to the transition temperature $T_c$ have been calculated using a model that takes into account effects of confinement and spontaneous chiral symmetry breaking ($\chi$SB), mainly focusing on the latter. 
Implications of $\chi$SB are realised in the form of dynamically generated quark masses as the temperature approaches $T_c$ from above. 
Through its induced kinematic constraint, this effect significantly suppresses the production of dileptons with large three-momenta. 
For dileptons emitted back-to-back with zero momentum the production rate is reduced more modestly.  

We have also examined the dependence of the dilepton production rate on the parameters that characterise the chiral crossover transition, suggesting that information about the transition can indeed be encoded in the production rate of dileptons with large three-momentum.
In particular variations of the transition temperature imply different spatial momenta at which the dilepton production vanishes as a consequence of phase space suppression.

\begin{figure}[t] 
\begin{center}
\includegraphics[width=0.50\textwidth]{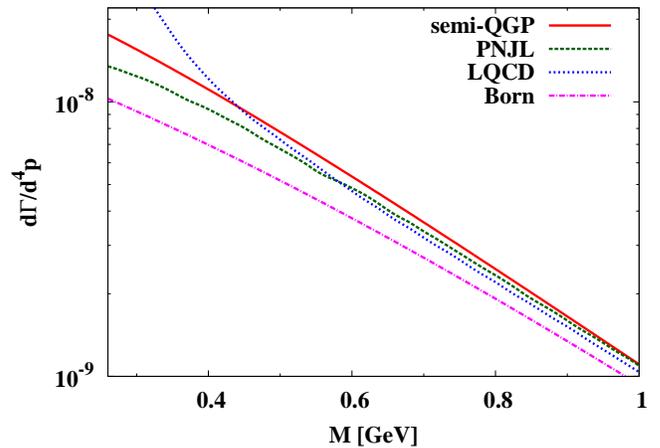} 
\caption{The dilepton production rate for $N_f = 2$ at $|\vp|=0$ and $T=240$ MeV as a function of the dilepton invariant mass $M$.
The solid (red) line represents our result (Eq.\,(\ref{eq:dilepton-result-p=0-N=3})), the dashed (green) line is the result of a PNJL model calculation~\cite{Islam:2014sea}, the dotted (blue) line is the result from lattice QCD~\cite{Ding:2010ga}. {{The dash-dotted (magenta) curve shows the perturbative (Born) dilepton rate (using Eq.\,(\ref{eq:dilepton-result-p=0-N=3}) with $A=B=C=1$)
}}.
} 
\label{fig:dileptonrate_comparison}
\end{center}
\end{figure}

A comparison of our calculated dilepton production rate with the results of other nonperturbative calculations is useful at this point.
The dilepton rate (Eq.~(\ref{eq:dilepton-result-p=0-N=3})) for $\Nf=2$ at $|\vp|=0$ and $T=240$ MeV is plotted as a function of $p^0$ in Fig.~\ref{fig:dileptonrate_comparison}.
The parameter set A was used to calculate $m(T)$.
Note that in this case $p^0=M\equiv \sqrt{p^2}$, the invariant mass of the dilepton,
and we use $M$ as the variable in this figure.
For comparison the results from a PNJL model calculation~\cite{Islam:2014sea}, lattice QCD~\cite{Ding:2010ga}, {{and a perturbative calculation are also shown.
The perturbative (Born approximation) result is obtained using Eq.~(\ref{eq:dilepton-result-p=0-N=3}) with $A=B=C=1$.}}
We note that our expression for the dilepton production rate (Eq.~(\ref{eq:dilepton-result-finitep})) {{has the same functional form in terms of $l$ and $m$ as}} that\footnote{From Eqs. (4.36) and (4.46) in Ref.~\cite{Islam:2014sea}, one can reproduce Eq.~(\ref{eq:dilepton-result-finitep}) in our paper by considering the case of zero chemical potential.
To confirm this, one needs to rewrite $f(E_p)-1$ in the two equations in Ref.~\cite{Islam:2014sea} to $-f(E_p-\omega)$ by changing the integration variable $p$.}
 obtained by using the PNJL model~\cite{Islam:2014sea}.
This is not surprising since the leading dilepton production amplitude ($q\overline{q}\rightarrow \gamma \rightarrow l\overline{l}$) is the same in both approaches.
Beyond leading order, loop corrections enter, which make the results depend on details of the model used.

At $T=240$ MeV the dynamical quark mass $m(T)$, in our approach as well as in the PNJL model, is negligible compared with $T$. The difference in Fig.~\ref{fig:dileptonrate_comparison} between our result and the PNJL model comes solely from the difference of the Polyakov loop values used in these calculations. 
Our smaller Polyakov loop stems from subtracting the perturbative correction, so that our dilepton production rate is enhanced compared to that of the PNJL model, as explained in Ref.~\cite{Gale:2014dfa, DileptonPhoton}.
{{The overall enhancement of the dilepton production rate induced by the Polyakov loop is also explicit in the fact that the perturbative (Born) rate is smaller than both the PNJL and semi-QGP results.}}

At small invariant mass $M$ one observes that our calculated dilepton rate is significantly lower than the one deduced from lattice QCD. One should note that the lattice results display considerable sensitivity
with respect to the ansatz chosen for the form of the vector current spectral function in the low-mass region, as
discussed in Ref.~\cite{Islam:2014sea}:
a ``free-field'' ansatz \cite{Karsch:2001uw} implies a very small dilepton production rate at small $M$, whereas
a ``Breit-Wigner'' ansatz \cite{Ding:2010ga} yields a large dilepton rate reflecting substantial contributions from
hadronic sources.

The photon production rate is found to be sensitive to $\chi$SB effects when $T$ is less than about 190 MeV.  We have studied the parameter dependence of this rate. It is found 
to vary significantly at low $T$ around $T_c$ depending on the choice of the transition temperature.

\begin{figure}[t] 
\begin{center}
\includegraphics[width=0.50\textwidth]{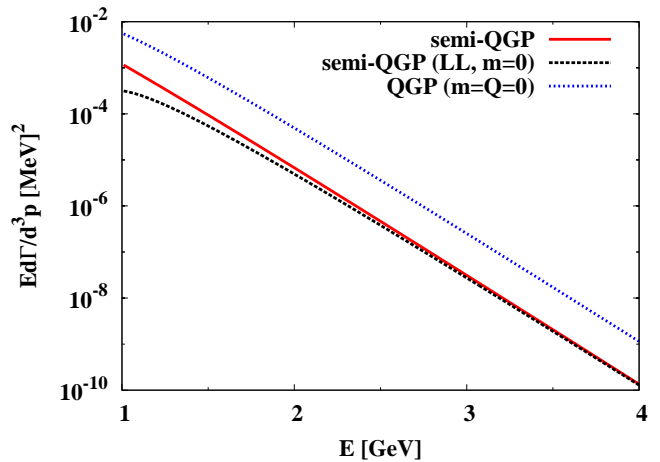} 
\caption{{The photon production rate for $\Nf=2$ at $T=180$ MeV as a function of the photon energy $E$.
Solid (red) line: present result (Eq.~(\ref{eq:photon-result})); dashed (black) line: semi-QGP model with $m=0$ and using leading-log approximation~\cite{Gale:2014dfa, DileptonPhoton}; dotted (blue) line (labeled QGP with $m = 0$ and $Q = 0$): perturbative QCD calculation including resummations~\cite{Baier:1991em, Kapusta:1991qp}. 
}} 
\label{fig:photonrate_comparison}
\end{center}
\end{figure}

{{Comparing our calculated photon rate with corresponding results using other methods is once again of some interest. Fig.~\ref{fig:photonrate_comparison} shows our result, Eq.~(\ref{eq:photon-result}), as a function of the photon energy $E$ in comparison with the semi-QGP photon rate~\cite{Gale:2014dfa, DileptonPhoton} taken at leading-log order and using $m=0$, and with the perturbative QGP result~\cite{Baier:1991em, Kapusta:1991qp}.
Here we have set $T=180$ MeV, $N_f =2$, and used the parameter set A.
The large-$\Nc$ limit is taken consistently in all three case studies. 
The semi-QGP approach at leading-log order and $m=0$ corresponds to Eq.~(\ref{eq:photon-result-m=0}) with 
\begin{align}
\label{eq:Z-LL}
Z=\pi^2\left(6 q^2-8 q+\frac{9}{4}\right)\ln\left(\frac{E}{g^2T}\right),
\end{align}
while the perturbative QGP result is given with $Z$ in Eq.~(\ref{eq:Z-result-largeN}) at $q=0$.
As seen in Fig.~\ref{fig:photonrate_comparison} the perturbative photon rate is larger than the semi-QGP rate over the whole energy range. The reason is that both Polyakov loop and constituent quark mass effects suppress the photon production as discussed in Sec.~\ref{ssc:photon-result} and Ref.~\cite{Gale:2014dfa, DileptonPhoton}.
At the lower energies $(E\sim 1-2$ GeV) our result exceeds the one calculated at leading-log order and with $m = 0$. This difference comes primarily from the term beyond leading-log. At high photon energies the leading-log form is approached asymptotically. Indeed, at sufficiently large $E$, the expressions for $X$ and $Y$ can be 
approximated by their leading-log forms, Eqs.~(\ref{eq:photon-compton-LL}) and (\ref{eq:photon-pair-LL}), so
that $X+Y$ equals $Z$ as given in Eq.~(\ref{eq:Z-LL}), except for the replacement $\ln (E/(g^2T))\rightarrow \ln(ET/m^2)$. At large $E$, this difference is negligible, and so the photon production rates in the two approaches become approximately equal.}}

A few concluding remarks are in order concerning possible $\chi$SB effects in high-energy heavy-ion collisions (HIC). Dilepton production in HIC is commonly discussed using the invariant mass $M$ instead of $|\vp|$. With inclusion of dynamical quark masses, the kinematical constraint for dilepton production reads $M > 2m(T)$. In the NJL model the dynamical quark mass is still quite large ($m \simeq 300$ MeV) at temperatures around 
$T\simeq 150$ MeV, so dileptons with $M< 600$ MeV are forbidden in this temperature range. At slightly higher temperatures where the dynamical quark mass decreases the kinematic constraint is correspondingly less stringent. 
While such effects can in principle be observed in HIC, one recalls, on the other hand, that the low-mass dilepton spectrum with $M \lesssim 0.8$ GeV results primarily from hadronic currents involving pseudoscalar and vector mesons rather than liberated quark-antiquark pairs \cite{Rapp:2000ff}. In this
range of invariant masses, dilepton spectra are governed by hadronic sources.

An interesting point of discussion is the photon elliptic flow $v_2$ observed at RHIC~\cite{PHENIX:2012}. As demonstrated in Ref.~\cite{Gale:2014dfa}, the suppression of the photon production rate from the QGP phase due to the confinement effect can increase the total elliptic flow, a welcome feature in order to understand the unexpectedly large $v_2$ for photons~\cite{Gale:2014dfa}.
Our results suggest that the effect of $\chi$SB can add to increasing the total photon $v_2$ further at $T$ between 150 and 190 MeV by its mechanism of suppressing the photon production in that temperature range. 
Of course, other sources of photon production in HIC also need to be considered, such as photons coming from the initial state~\cite{Aurenche:2006vj}, from the thermalization process~\cite{McLerran:2014hza}, and from the hadron phase~\cite{Turbide:2003si}. A complete evaluation of the photon $v_2$ requires summing the contributions from all the photon sources in extended hydrodynamic simulations~\cite{Gale:2014dfa, Schenke:2010nt, Schenke:2010rr}.

\section*{Acknowledgements}
The authors thank Robert Lang for providing numerical results of the temperature-dependent NJL constituent quark mass. 

\appendix*
\section{Photon Production Rate when $m=0$}
\label{app:photon-massless}

\begin{figure}[t] 
\begin{center} 
\includegraphics[width=0.15\textwidth]{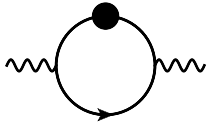} 
\caption{The photon self-energy at the one-loop level. 
The solid line with a blob represents the HTL-resummed quark propagator.
} 
\label{fig:photon-HTLresum}
\end{center}
\end{figure}

In this appendix we derive the photon production beyond leading-log order in the limit of vanishing quark mass, $m = 0$.
The result at leading-log order has been reported previously~\cite{Gale:2014dfa, DileptonPhoton}, but the derivation beyond this order has not been performed yet.
The infrared singularities in the contributions from Compton scattering and pair annihilation require introducing a cutoff $\mu$ as a regulator. In principle
$\mu$ is an arbitrary quantity, but it should satisfy $gT\ll \mu\ll T$.
The phase space integral appearing in the Compton scattering process and the pair annihilation contributions are restricted by the cutoff $\mu$~\cite{Kapusta:1991qp}:
$2\mu^2<s$ and $-s+\mu^2<t<-\mu^2$.
In the phase space excluded by $\mu$, the momentum of the exchanged quark is smaller than $\mu$, and thus much smaller than $T$.
Such soft quarks should be treated using the hard thermal loop (HTL) resummed propagator~\cite{Hidaka:2009hs, Pisarski:1988vd}, which contains the information of the thermal mass of the quark, so we calculate the resummed one-loop diagram drawn in Fig.~\ref{fig:photon-HTLresum} in order to obtain the photon production rate due to the soft momentum exchange.

The hard parts from Compton scattering and pair annihilation are given by Eqs.~(\ref{eq:Compton-expression}) and (\ref{eq:pair-expression}), setting $m=0$, and the phase space integration is constrained by the cutoff $\mu$ as explained above.
The resulting expressions are as follows~\cite{DileptonPhoton}:
\begin{align} 
\nonumber
\left. E\frac{d\varGamma}{d^3\vp}\right|_{\text{Comp}}
&\simeq  -\frac{\alpha\alpha_s}{8\pi^4}\frac{T}{E} e^{-\beta E}\sum_{f,a\sim d} q^2_f 
\int^\infty_{2\mu^2} d{s} \int^{-\mu^2}_{-s+\mu^2} d{t} \\
\nonumber
&~~~\times \frac{1}{{s}} 
 e^{i\beta(Q_a-Q_b+Q_c-Q_d)} [(t^{cd})_{ba}]^2 \\
&~~~\times  \ln\left(1+e^{-\beta({s}/(4E)-iQ_b)}\right) 
\left(\frac{{t}}{s} +\frac{{s}}{t} \right),\\
\nonumber
\left. E\frac{d\varGamma}{d^3\vp}\right|_{\text{pair}}
&\simeq -\frac{\alpha\alpha_s}{16\pi^4}\frac{T}{E}e^{-\beta E}\sum_{f,a\sim d}q^2_f  
\int^\infty_{2\mu^2} d{s} \int^{-\mu^2}_{-s+\mu^2} d{t} \\
\nonumber
&~~~\times \frac{1}{s}  e^{i\beta(Q_a-Q_b-Q_c+Q_d)}   [(t^{dc})_{ba}]^2\\ 
   &~~~\times \ln\left(1-e^{-\beta s/(4E)}e^{i\beta(Q_c-Q_d)}\right)
 \left(\frac{u}{{t}} +\frac{t}{{u}}\right).
\end{align}
Here we have used the Boltzmann approximation for the initial state.
By performing these integrations, we get
\begin{align}  
\nonumber
\left. E\frac{d\varGamma}{d^3\vp}\right|_{\text{Comp}}
&\simeq  \frac{\alpha\alpha_s}{4\pi^4}T^2 e^{-\beta E}\sum_{f} q^2_f (\Nc^2-1)
\sum^\infty_{n=1} \frac{(-1)^{n+1}}{n^2}\\
&~~~\times l_n
\left[\ln\left(\frac{4ET}{\mu^2}\right)+\frac{1}{2}-\gamma_E-\ln n\right],\\
\nonumber
\left. E\frac{d\varGamma}{d^3\vp}\right|_{\text{pair}}
&\simeq  \frac{\alpha\alpha_s}{4\pi^4}T^2 e^{-\beta E}\sum_{f} q^2_f 
\sum^\infty_{n=1} \frac{1}{n^2}\left[(\Nc l_n)^2-1\right] \\
&~~~\times \left[\ln\left(\frac{4ET}{\mu^2}\right)-1-\gamma_E-\ln n\right].
\end{align}
In the large-$\Nc$ limit these expressions become
\begin{align}  
\nonumber
\left. E\frac{d\varGamma}{d^3\vp}\right|_{\text{Comp}}
&\simeq  \frac{\alpha\alpha_s}{4\pi^4}T^2 e^{-\beta E}\sum_{f} q^2_f \Nc^2
\sum^\infty_{n=1} \frac{(-1)^{n+1}}{n^2}l_n \\
&~~~\times \left[\ln\left(\frac{4ET}{\mu^2}\right)+\frac{1}{2}-\gamma_E-\ln n\right],\\
\nonumber
\left. E\frac{d\varGamma}{d^3\vp}\right|_{\text{pair}}
&\simeq  \frac{\alpha\alpha_s}{4\pi^4}T^2 e^{-\beta E}\sum_{f} q^2_f \Nc^2
\sum^\infty_{n=1} \frac{1}{n^2}(l_n)^2\\
&~~~\times \left[\ln\left(\frac{4ET}{\mu^2}\right)-1-\gamma_E-\ln n\right],
\end{align}

With $\Nc=3$ they reduce to:
\begin{align}  
\nonumber
\left. E\frac{d\varGamma}{d^3\vp}\right|_{\text{Comp}}
&\simeq  \frac{\alpha\alpha_s}{4\pi^4}T^2 e^{-\beta E}\sum_{f} q^2_f 
3\\
\nonumber
&~~~\times\Biggl[\left(\ln\left(\frac{4ET}{\mu^2}\right)+\frac{1}{2}-\gamma_E\right)\\
\nonumber
&~~~\times\frac{\pi^2}{4}(1-8q^2)
+0.10132 \\ 
\label{eq:photon-compton-largeN-N=3}
&~~~-2\sum^\infty_{n=1} \frac{(-1)^{n+1}}{n^2}\cos(2\pi nq)\ln n\Biggr],\\
\nonumber
\left. E\frac{d\varGamma}{d^3\vp}\right|_{\text{pair}}
&\simeq  \frac{\alpha\alpha_s}{4\pi^4}T^2 e^{-\beta E}\sum_{f} q^2_f \\
\nonumber
&~~~\times\Bigl[\left(\ln\left(\frac{4ET}{\mu^2}\right)-1-\gamma_E\right)\frac{3\pi^2}{2}\\
\nonumber
&~~~\times\left(8q^2-\frac{16}{3}q+1\right)\\
\nonumber
&~~~-2.8127
-2\sum^\infty_{n=1} \frac{1}{n^2}\\
\label{eq:photon-pair-largeN-N=3}
&~~~\times\left(2\cos(2\pi nq)+\cos(4\pi nq)\right)\ln n\Bigr],
\end{align}
by using Eq.~(\ref{eq:ln-Q}), and 
\begin{align} 
\sum^\infty_{n=1} \frac{(-1)^{n+1}}{n^2}\cos(2\pi nq)
&= -\pi^2\left(q^2-\frac{1}{12}\right),\\
\sum^\infty_{n=1} \frac{1}{n^2} \cos(2\pi nq)
&= \pi^2\left(q^2-|q|+\frac{1}{6}\right),\\
-\sum^\infty_{n=1} \frac{(-1)^{n+1}}{n^2}\ln n
&= \frac{1}{12}\left(\pi^2\ln 2+6\zeta'(2)\right) \\
\nonumber
&\simeq 0.10132,\\ 
\nonumber
\sum^\infty_{n=1} \frac{1}{n^2} \ln n
&= -\frac{\pi^2}{6}
(\gamma_E+\ln 2\pi \\
&~~~-12\ln C_{\text{Glaisher}})
\simeq 0.93755,
\end{align}
where $C_{\text{Glaisher}}\simeq 1.2824$ is the Glaisher constant.

On the other hand, the contribution from the soft part reads~\cite{DileptonPhoton}
\begin{align} 
\begin{split}
\left. E\frac{d\varGamma}{d^3\vp}\right|_{\text{soft}}
&= \frac{2e^2}{(2\pi)^3} \nb(E) 
\sum_f q^2_f \sum_a\int\frac{d^3\vk}{(2\pi)^3}
\int\frac{d\omega}{2\pi} \\
&~~~\times\left[\nf{}_{a}(\omega-E)-\nf{}_{a}(\omega)\right]\\
&~~~\times\Tr\left[ \rho^*_a(\omega,\vk) \rho(\omega-E,\vk-\vp) \right],
\end{split}
\end{align} 
which corresponds to the one-loop diagram drawn in Fig.~\ref{fig:photon-HTLresum}.
The summation includes the color indices $a$. Here the bare (HTL-resummed) quark spectral functions are:
\begin{align}
\label{eq:propagator-spectral-bare}
\rho(\omega,\vk)&=2\pi\varepsilon(\omega)\,\Slash{k}\, \delta(k^2),\\
\label{eq:propagator-spectral-HTL}
\rho^*_a(\omega,\vk)&= 
\frac{\gamma^0-\vgamma\cdot\hat{\vk}}{2}\rho^*_{a+}(\omega,\vk)
+ \frac{\gamma^0+\vgamma\cdot\hat{\vk}}{2}\rho^*_{a-}(\omega,\vk),
\end{align}
where $\varepsilon(\omega)$ is the sign function.
The HTL spectral function after decomposition reads
\begin{align}
\label{eq:HTL-spectral}
\begin{split}
\rho^*_{a\pm}(\omega,\vk)&= 2\pi [Z_{\pm a}(|\vk|)  \delta(\omega-\omega_{\pm a}(|\vk|)) \\
&~~~+ Z_{\mp a}(|\vk|)  \delta(\omega+\omega_{\mp a}(|\vk|)) ]\\
&~~~+\theta(|\vk|^2-\omega^2)\rho^{\text{spacelike}}_{a\pm}(\omega,\vk)
\end{split}
\end{align}
Here $Z_{\pm a}(|\vk|)$ is the residue, $\omega_{\pm a}(|\vk|)$ the dispersion relation, and $m_a(T)$ is the thermal quark mass. Their respective expressions are:
\begin{align}
Z_{\pm a}(|\vk|)&=\frac{\omega^2_{\pm a}(|\vk|)-|\vk|^2}{2\, m_a},\\
\nonumber
\omega_{\pm a}(|\vk|)\mp &|\vk|\\
= &\frac{\mf^2{}_a}{|\vk|}
\left[\left(1\mp\frac{\omega_{\pm a}(|\vk|)}{|\vk|}\right)Q_0\left(\frac{\omega_{\pm a}(|\vk|)}{|\vk|}\right)
\pm 1\right],
\end{align}
where $Q_0(x)\equiv (\ln(x+1)/(x-1))/2$.
The explicit form of $\rho^{\text{spacelike}}_{a\pm}$ is irrelevant in our analysis, so we do not write it here.
The thermal quark mass is given by~\cite{Hidaka:2009hs}
\begin{align}
\label{eq:thermalmass-semiQGP}
\begin{split}
m^2_{a}&=\frac{g^2T^2}{4}
\Biggl[-\sum^{\Nc}_{b=1}|q_a-q_b|+\Nc\left(q^2_a+\frac{1}{4}\right)^2 \\
&~~~+\frac{1}{\Nc}\left(q^2_a-\frac{1}{4}\right)^2\Biggr] ,
\end{split}
\end{align}
where $q_a\equiv Q_a/(2\pi T)$.
At large $\Nc$, the last term is neglected.
The production rate is reduced to the following form~\cite{DileptonPhoton}:
\begin{align} 
\begin{split}
\left. E\frac{d\varGamma}{d^3\vp}\right|_{\text{soft}}
&= \frac{4\alpha}{(2\pi)^3} e^{-\beta E}  
\sum_f q^2_f \sum_a\left(\nf{}_a(0)-1\right) m^2_{a} \\
&~~~\times\Biggl[-1+\ln\frac{\mu^2}{m^2_{a}}
+\frac{2}{m^2_a}\int^\infty_0 d|\vk| \\
&~~~\times \left(\omega_{+a}(|\vk|)-\omega_{-a}(|\vk|)-\frac{m^2_a}{|\vk|+m_a}\right) 
\Biggr].
\end{split}
\end{align} 
The integration in the equation above does not depend on $a$, and its numerical value approximately $ -0.16258$~\cite{Baier:1991em}.

Using Eq.~(\ref{eq:thermalmass-semiQGP}) in the large-$\Nc$ limit, and considering the case of $\Nc=3$, we get
\begin{align}
m^2_{1}=m^2_{3}&= \frac{3g^2T^2}{16}\left(1-4q+4q^2\right),\\
m^2_{2}&= \frac{3g^2T^2}{16}\left(1-\frac{8}{3}q\right).
\end{align}
With these expressions one finds:
\begin{widetext}
\begin{align} 
\label{eq:photon-soft-largeN-N=3}
\begin{split}
\left. E\frac{d\varGamma}{d^3\vp}\right|_{\text{soft}}
&=\frac{\alpha\alpha_s}{4\pi^4}T^2 e^{-\beta E}\sum_{f} q^2_f 
\frac{3\pi^2}{2} 
\Biggl[\left(\ln\left(\frac{\mu^2}{g^2T^2}\right) -1.32516\right)
\frac{3}{2}\left(\frac{8}{3}q^2-\frac{32}{9}q+1\right) \\
&~~~-\left(1-4q+4q^2\right)\ln \frac{3}{16}\left(1-4q+4q^2\right) 
-\frac{1}{2}\left(1-\frac{8}{3}q\right)\ln \frac{3}{16}\left(1-\frac{8}{3}q\right)
\Biggr]
\end{split}
\end{align}
\end{widetext}
The final result in the large $\Nc$ limit is given by the sum of the Compton, pair annihilation, and soft parts, namely Eqs.~(\ref{eq:photon-compton-largeN-N=3}), (\ref{eq:photon-pair-largeN-N=3}), and (\ref{eq:photon-soft-largeN-N=3}), when $\Nc=3$.


\end{document}